\documentclass[sigconf]{acmart}
\usepackage{graphicx}
\usepackage{stfloats}
\usepackage{array}
\usepackage{algorithmic}
\usepackage{algorithm}
\usepackage{natbib}
\usepackage{xcolor}
\usepackage{svg}
\usepackage{subcaption}

\AtBeginDocument{%
  \providecommand\BibTeX{{%
    \normalfont B\kern-0.5em{\scshape i\kern-0.25em b}\kern-0.8em\TeX}}}

\copyrightyear{2023}
\acmYear{2023}
\setcopyright{rightsretained}
\acmConference[SC-W 2023]{Workshops of The International Conference on High Performance Computing, Network, Storage, and Analysis}{November 12--17, 2023}{Denver, CO, USA}
\acmBooktitle{Workshops of The International Conference on High Performance Computing, Network, Storage, and Analysis (SC-W 2023), November 12--17, 2023, Denver, CO, USA}
\acmDOI{10.1145/3624062.3624228}
\acmISBN{979-8-4007-0785-8/23/11}

\begin{document}

\title{RISA: Round-Robin Intra-Rack Friendly Scheduling Algorithm for Disaggregated Datacenters}

\author{Rashadul Kabir}
\affiliation{%
 \institution{Colorado State University}
 \city{Fort Collins}
 \state{CO}
 \country{USA}}
\email{rashadul.kabir@colostate.edu}

 \author{Ryan G. Kim}
\affiliation{%
 \institution{Intel Labs}
 \city{Hillsboro}
 \state{OR}
 \country{USA}}
 \email{ryan.gary.kim@intel.com}

 \author{Mahdi Nikdast}
\affiliation{%
 \institution{Colorado State University}
 \city{Fort Collins}
 \state{CO}
 \country{USA}}
 \email{mahdi.nikdast@colostate.edu}
\renewcommand{\shortauthors}{Kabir et al.}

\begin{abstract}
Recent trends see a move away from a fixed-resource server-centric datacenter model to a more adaptable ``disaggregated'' datacenter model. These disaggregated datacenters can then dynamically group resources to the specific requirements of an incoming workload, thereby improving efficiency. To properly utilize these disaggregated datacenters, workload allocation techniques must examine the current state of the datacenter and choose resources that not only optimize the current workload request, but future ones. Since disaggregated datacenters are severely bottlenecked by the available network resources, our work proposes a heuristic-based approach called RISA, which significantly reduces the network usage of workload allocations in disaggregated datacenters. Compared to the state-of-the-art, RISA reduces the power consumption for optical components by $33\%$ and reduces the average CPU-RAM round-trip latency by $50\%$. Additionally, RISA significantly outperforms the state-of-the-art in terms of execution time.
\end{abstract}


\maketitle

\section{Introduction}\label{sec1}
\begin{figure}[ht]
    \centering
    \includegraphics[width=\linewidth]{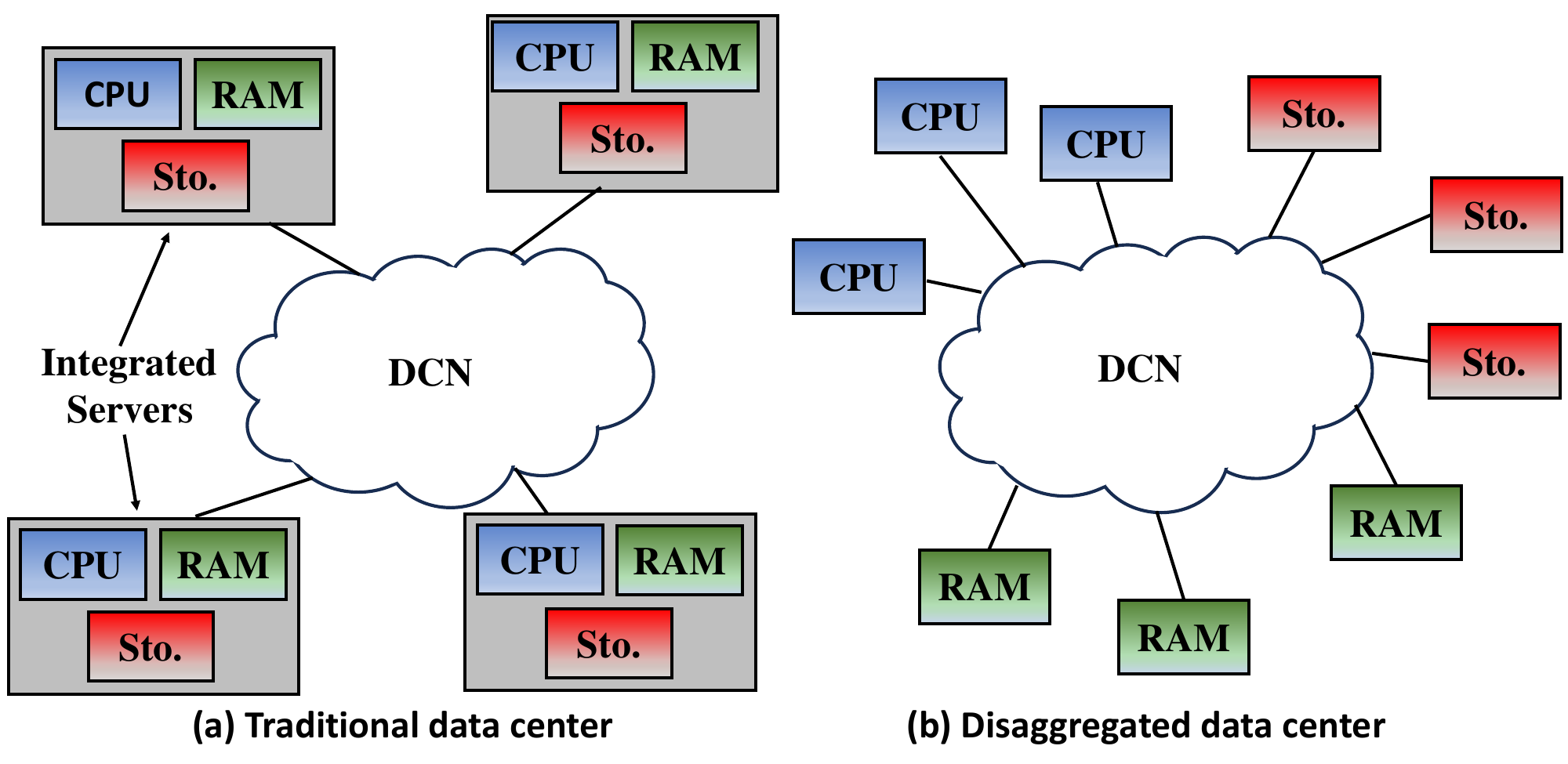} 
    \caption{Disaggregated vs. Traditional}
    \label{ddc0}
\end{figure}

Traditionally, datacenters~(DCs) have been server-centric, characterized by network-connected homogeneous servers with fixed ratios of compute, memory, and storage. Modern cloud workloads, however, demand diverse resource ratios, leading to inefficiencies and significant amounts of stranded resources. These unused stranded resources not only increase the capital costs but also amplify power consumption, costing up to 85\% of total DC expenses~\cite{Barroso}. Furthermore, although the life cycle and technological advancements of various server resources differ, this fixed integration paradigm requires any hardware upgrade or resource expansion to be executed at the server level~\cite{Guo}. Figure~\ref{ddc0} shows how a disaggregated DC (DDC) is different from a traditional datacenter. 

In terms of supporting the implementation of DDC systems, researchers have proposed an alternative to the standard network interface card, known as a switch and interface card (SIC). SIC is able to perform the packet and circuit switching services required in a DDC~\cite{Yan}. Researchers have also investigated operating systems focusing on resource disaggregation~\cite{OS}. Beyond the academic realm, the industry too has shown a keen interest, with notable contributions including the Intel Rack Scale Architecture~\cite{Intel}, dReDBox project~\cite{dREdBox}, and Firebox~\cite{Firebox}. Throughout these advancements, both in academia and industry, the central aim has been to maximize resource utilization.

Despite the enthusiasm surrounding resource disaggregation, it is challenging to design the prerequisite network infrastructure. In order for disaggregated datacenters (DDCs) to be desirable, the network should achieve similar latency and bandwidths to their traditional direct-attached counterparts while keeping costs and power consumption low. Furthermore, a coordinated orchestration of compute, memory, and network resources is essential to maximize resource utilization and workload performance, while simultaneously keeping both latency and costs low. Thus, compute and network scheduling becomes an integral part of research in resource disaggregation.

One of the seminal efforts in disaggregated resource scheduling was done by Zervas~\textit{et al.}~\cite{Zervas}. In~\cite{Zervas}, the authors propose two algorithms for scheduling virtual machines~(VMs) onto disaggregated CPU, RAM, and storage nodes to maximize resource utilization. The first is a network-unaware locality-based~(NULB) heuristic-based scheduling algorithm that uses a breadth-first search~(BFS) to choose resources. The second algorithm is a network-aware locality-based~(NALB) scheduling algorithm that extends NULB to also consider network utilization. NALB chooses to use links with higher available bandwidth; thus, tries to ensure that VMs are not dropped because of unavailable link bandwidth. Later, in Section~\ref{risa_overview}, we will demonstrate that NULB and NALB suboptimally utilize network resources. In particular, they utilize more inter-rack network resources, which motivates the need for an approach that focuses on optimally utilizing both the compute and network resources. Thus, in this paper, we propose a novel heuristic-based algorithm called Round-robin Intra-rack friendly Scheduling Algorithm~(RISA), which is able to schedule workloads onto disaggregated CPU, RAM, and storage nodes while attempting to maximize resource utilization and minimize network utilization and CPU-RAM latency. 

The rest of the paper is organized as follows. Section~\ref{related_work}, discusses some developments in resource disaggregation and workload scheduling in DDC architectures. Section~\ref{background} discusses the DDC architecture used in this paper and the optical switch energy model. Section~\ref{risa_overview} details our proposed approaches RISA and RISA-BF, including deficiencies in prior work. Section~\ref{sim_res}, presents our simulation results and comparative analysis. Finally, concluding remarks are given in Section~\ref{conc}.

\section{Related work}\label{related_work}
Over the years, there have been efforts to separate or disaggregate server resources. Significant advancements have been achieved through the use of Storage Area Networks~(SANs) and Network-Attached Storage~(NAS) systems, both of which offer storage solutions over a network~\cite{Zervas}. In 2009, memory disaggregation was introduced to address memory capacity challenges~\cite{Lim}. Following a short period of diminished activity, the field of resource disaggregation experienced a resurgence in 2016. The first work on workload scheduling in DDCs was by Papaioannou~\textit{et al.}~\cite{Benefits}. They proposed a heuristic for scheduling VMs onto rack-scale DDC, showcasing how resource utilization can be better compared to traditional DC scheduling techniques. They considered only CPU and RAM in their scheduling problem. Another critical difference is the consideration of inter-VM communication, their approach additionally attempted to schedule VMs that communicate with one another closer. However, for DCs that primarily service third-party workloads, VMs may function independently. One such example is the Azure data center traces~\cite{azure2023}. Ali~\textit{et al.}~\cite{Ali} proposed an MILP-based energy-aware scheduling approach at the DC scale. They considered CPU, RAM, and IO resources. Next, Zervas~\textit{et al.}~\cite{Zervas} proposed the NULB and NALB algorithms. These algorithms consider CPU, RAM, and storage for the scheduling problem. The primary focus of this paper was to minimize the number of dropped VMs and maximize resource utilization. However, the manner in which compute resource search was prioritized in these algorithms, it encouraged inter-rack VM assignments. One good aspect of~\cite{Zervas}, was the DDC architecture used in the paper. It was heavily inspired by the DDC architecture developed by IBM, dRedBox~\cite{dREdBox}. As a follow-up to~\cite{Zervas},~\cite{Shabka} proposed a reinforcement learning-based algorithm, focusing on reduced network usage. Although~\cite{Shabka} significantly outperformed NULB and NALB~\cite{Zervas}, the problem definition is different from that in our work. First, the datacenter network in~\cite{Shabka} uses a three-tier tree network structure while our work uses only two tiers. The three-tier structure scheduling problem requires consideration of intra-rack and inter-rack within the same sub-tree, and inter-rack among different sub-tree resource relationships. Our focus is to schedule resources within a cluster (intra-rack and inter-rack), limiting the maximum number of hops. Second,~\cite{Shabka} allows VMs that require more than one box of the same resource type. In our work, the VM resource requirements are always smaller than the capacity of one resource box. Third, storage is not considered in~\cite{Shabka}. These last two points significantly change the nature of the VM requests and especially the network considerations needed during scheduling. Additionally, we focus on developing a deterministic heuristic to schedule compute and network resources in a disaggregated datacenter. Since~\cite{Shabka} is a non-deterministic machine learning-based approach, it is beyond the scope of this work. Other researchers~\cite{Chao, Call} have approached the DDC workload scheduling problem from different angles. However, none of them considered all three resource types (CPU, RAM, and storage) for the scheduling problem, while utilizing an industry-standard DDC architecture.



\section{Background}\label{background}
\subsection{Disaggregated architecture}
\begin{figure}[ht]
    \centering
    \includegraphics[width=\linewidth]{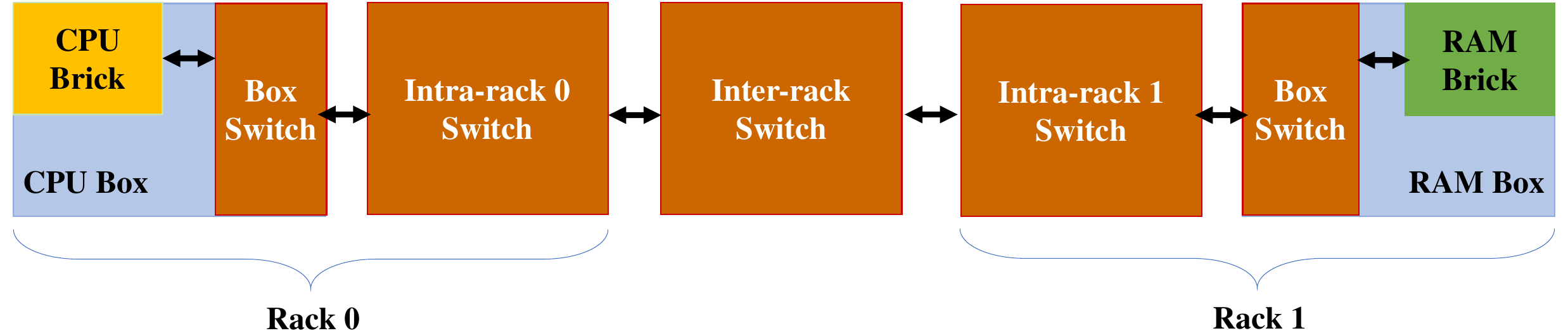} 
    \caption{Inter-rack communication in the disaggregated architecture proposed in~\cite{Zervas}}
    \label{inter_rack_comm}
\end{figure}

\begin{figure*}[ht]
    \centering
    \includegraphics[width=\textwidth]{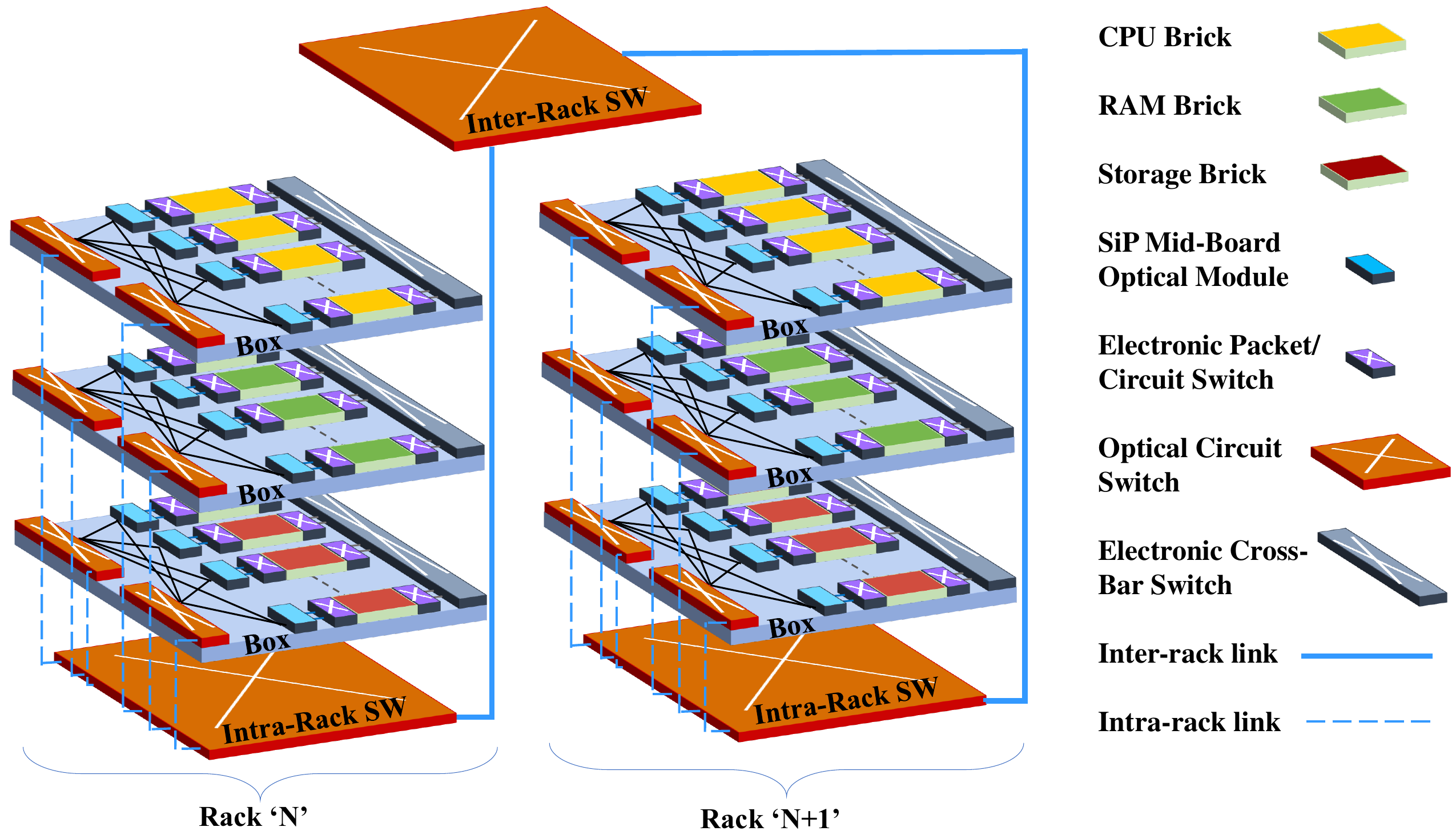} 
    \caption{The disaggregated datacenter architecture proposed in~\cite{Zervas} used as the case study in this work }
    \label{ddc2}
\end{figure*}

The disaggregated datacenter architecture proposed in~\cite{Zervas} is used as the case study in this work (see Figure~\ref{ddc2}). This architecture is structured into racks. Within each rack, there are several boxes, each box has some amount of a single resource type: CPU, RAM, or storage. These boxes are further divided into bricks, with each brick holding a predetermined quantity of its designated resource. Each box is equipped with optical switches, connected to rack-level switches, for communication. Progressing up the hierarchy, rack switches then connect to inter-rack switches. As seen in Figure~\ref{inter_rack_comm}, if a CPU brick within a CPU box in rack 0 intends to interact with a RAM brick within a RAM box in rack 1, the communication journey would entail traversing the box switch of rack 0, its rack switch, the inter-rack switch, the switch of rack 1, and finally, the RAM box's switch before reaching its destination. 

Table~\ref{conf} offers a detailed configuration utilized in our simulation study. Notably, within a brick, all communication is electronic. Once the data leaves the brick, it gets converted from electronic to photonic by a single-mode Luxtera commercial SiP optical module~\cite{luxtera2023} (depicted in Figure~\ref{ddc2}). This module uses eight spatially multiplexed optical channels, each capable of supporting a 25Gb/s bit rate, for a total of 200Gb/s per link. Utilizing transceivers that operate in single-mode, rather than multimode, enables a more scalable network built on optical circuit switching (OCS). This is because single-mode fibers accommodate optical switches with a greater number of ports~\cite{Nagashima}. We obtain the power consumption of the transceiver module to be 22.5~$pJ/bit$ from~\cite{Zervas}. Using Tables~\ref{conf} and \ref{net_req} \cite{Zervas}, we can get a sense of network requirements for different sizes of VMs. 

\begin{table}[ht!]
\centering
\caption{Disaggregated architecture configuration}
\begin{tabular}{|l l|l l|}
\hline
\multicolumn{2}{|c|}{DDC Configuration} & Brick size & 16 units \\
\hline
Cluster size & 18 racks & CPU unit & 4 cores \\
\hline
Rack size & 6 boxes & RAM unit & 4 GB \\
\hline
Box size & 8 bricks & Storage unit & 64 GB \\
\hline
\end{tabular}
\label{conf}
\end{table}

\begin{table}[ht!]
\centering
\caption{Network requirements}
\begin{tabular}{|l l|l l|}
\hline
\multicolumn{2}{|c|}{CPU-RAM bandwidth} & \multicolumn{2}{|c|}{5 Gb/s/unit}\\
\hline
\multicolumn{2}{|c|}{RAM-STO bandwidth} & \multicolumn{2}{|c|}{1 Gb/s/unit}\\
\hline
\end{tabular}
\label{net_req}
\end{table}

\subsection{Optical switch energy model}

To accurately model the optical switch power, we need to consider the states of the cells within a switch. Known for their fast reconfiguration times, we consider microring resonator (MRR)-based switches for our disaggregated setup. For the switch configuration, we selected a commonly used Bene{\v s} network configuration. 

Within an optical switch, ports are interconnected through a network of cells. Each cell in the switch can be in either ``cross'' or ``bar'' state. When a VM requests a path through a switch, some cells must change their states. For this, each of these cells consumes switching power ($P_{swcell}$). Thus, if there are $n$ number of cells along a switch path, we assume that $n/2$ of the cells will undergo reconfiguration. For the rest of the time, to maintain a cell's state it consumes trimming power ($P_{trimcell}$). Therefore, the optical switch energy consumed for each VM ($E_{sw}$) is:

\begin{equation}
\label{eq:swpower}
\begin{split}
E_{sw} = (\frac{n}{2} \times P_{swcell} \times lat_{sw}) + (\alpha \times n \times P_{trimcell} \times T)
\end{split}
\end{equation}
where $lat_{sw}$ is the latency of switching the cells (dependent on the switch size~\cite{De}), $\alpha$ is a constant to consider the fact that two VMs can share the same cell, and $T$ is the lifetime of the VM. 

Based on the findings in~\cite{Mirza}, we considered $P_{trimcell} = 22.67 mW$ and $P_{swcell} = 13.75 mW$.  The number of cells in a Bene{\v s}-based optical switch is dependent on the number of ports on the switch, as detailed in~\cite{Lee}. Similarly, the latency of switching the cells are also based on the switch size~\cite{De}. The switch configurations are detailed in the experimental results in Section~\ref{sim_res}. Since two VMs can share the same switch cell $\alpha$ should be between $0.5$ (every cell is shared) and $1$ (no cell is shared). For the purpose of our simulations, we've chosen the value of $\alpha$ to be $0.9$.

\begin{figure}[ht]
    \centering
    \includegraphics[width=\linewidth]{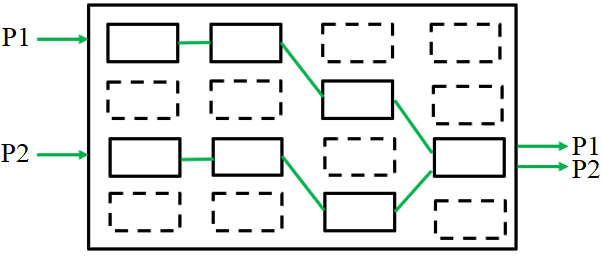} 
    \caption{Generic view of an optical switch}
    \label{ocs1}
\end{figure}

Figure~\ref{ocs1} shows a generic view of an optical switch, represented by the largest box. The boxes inside the switch with solid outlines represent the cells that are being utilized (dashed are inactive). In the figure, we show two paths, $P1$ and $P2$, corresponding to two VMs that share an MRR cell. 

\section{RISA Overview}\label{risa_overview}
\subsection{Discussion of NULB and NALB}
Before discussing RISA in detail, it's crucial to first grasp the workings and limitations of NULB and NALB. Both of these algorithms have a compute~(CPU/RAM/storage) resource allocation phase and a network resource allocation phase. 

During the compute resource allocation phase, \cite{Zervas} prioritizes the most scarce resource by examining the contention ratio~(CR) or the amount of a resource required by a VM over the total amount of that available resource. NULB and NALB both first search for a suitable box to satisfy the resource request with the highest CR. Then, NULB searches for the other resources using breadth-first search (BFS), while NALB uses a modified BFS. In modified BFS, NALB reorders neighbors of the scarce resource in descending order of their available bandwidth. In the network resource allocation phase, NULB selects the first available link to establish the connection between each pair of resources. NALB chooses links with the most available bandwidth. 

The way the compute resource search is prioritized in NULB or NALB, it encourages inter-rack network utilization. Thus, even though the network allocation phase of NALB is network-aware, it does not truly discourage higher network utilization.

\subsection{Discussion of RISA}

To better assign VMs to intra-rack resources, RISA keeps track of the boxes with the maximum amount of each resource for each rack. Thus, when a VM requests resources, we can easily determine which racks have boxes that can accommodate the entire VM. These racks that can house the entire VM are placed in a list called $INTRA\_RACK\_POOL$. To help balance the load between racks, we adopt a round-robin policy for selecting racks from $INTRA\_RACK\_POOL$. This helps to make the utilization of the racks more uniform. Algorithm~\ref{RISA} details the different steps of RISA. $REQ$ indicates the VM compute requirements. $NET$ indicates the total network bandwidth available. $AVAIL\_INTRA\_RACK\_NET$ indicates the total intra-rack network bandwidth available.

If \textit{INTRA\_RACK\_POOL} is empty or available intra-rack network resources is insufficient to schedule a VM using \textit{INTRA\_RACK\_POOL}, we create three lists, known as the $SUPER\_RACK$ collectively. Within the $SUPER\_RACK$, each list contains the racks with boxes with sufficient resources (CPU or RAM or storage) for a VM assignment. In such a scenario, the VM assignment has to be inter-rack. In this case, RISA resorts to NULB~\cite{Zervas} (Algorithm~\ref{NULB}) to perform the VM allocation. The choice boxes for VM assignment are then limited to the boxes of racks within the SUPER\_RACK. Also, VM assignment starts with first finding the most contended resource, then based on BFS, other resources are first searched for in the same rack and then searched for in other racks.

\begin{algorithm}[ht!]
\caption{RISA}
\begin{algorithmic}
\WHILE{$\exists$ unscheduled VM}
    \STATE Create $INTRA\_RACK\_POOL$
    \IF{$INTRA\_RACK\_POOL \neq \emptyset$}
        \FORALL{$rack \in INTRA\_RACK\_POOL$ (round-robin)}
            \IF{$AVAIL\_INTRA\_RACK\_NET \neq \emptyset$}
                \STATE $Compute \leftarrow AllocCom(INTRA\_RACK\_POOL, REQ)$
                \STATE $Network \leftarrow AllocNet(AVAIL\_INTRA\_RACK\_NET)$
                \IF{$(Compute \neq \emptyset) and (Network \neq \emptyset)$}
                \STATE Assign VM
                \ENDIF
            \ENDIF
        \ENDFOR
    \ELSE
        \STATE Create $SUPER\_RACK$
    \STATE $Compute, Network \leftarrow  NULB(SUPER\_RACK, REQ, NET)$
        \IF{($Compute \neq \emptyset$) \AND ($Network \neq \emptyset$)}
            \STATE Assign VM
        \ENDIF
    \ENDIF
\ENDWHILE
\end{algorithmic}
\label{RISA}
\end{algorithm}

Algorithm~\ref{NULB} outlines the different steps of NULB. $res\_type$ indicates all kinds of resources (e.g., CPU, RAM, and storage). After finding the most scarce resource, $res_{max}$ using CR, NULB first looks for the first box with the available $res_{max}$ requested by a VM. Next, using BFS, NALB first looks for other requested resources by VM in the same rack. If it doesn't find them in the same rack it looks for resources in other racks. Once a set of boxes with available resources is found, this indicates successful completion of the compute allocation phase. Next, NALB checks to see if the available network bandwidth is present in the connecting optical links. If it finds the links with the necessary bandwidth, the VM assignment is successful with the completion of the network allocation phase. If either the compute allocation or network allocation fails, the VM to be assigned is dropped.

\begin{algorithm}[ht!]
\caption{NULB}
\begin{algorithmic}
\STATE $NULB(RES, REQ, NET)$
    \FORALL{$res\_type$}
        \STATE Append $CR(res\_type)$ to $CR\_LIST$
    \ENDFOR
    \STATE $res_{max}$ $\leftarrow$ $max(CR\_LIST)$
    \IF{$res_{max}$ is available on any box}
        \STATE $Compute \leftarrow AllocCompute(BFS(res_{max}), REQ)$
        \IF{$Compute \neq \emptyset$}
            \STATE $Network \leftarrow AllocNetwork(NET)$
            \IF{$Network \neq \emptyset$}
                \STATE Assign VM (when implementing NULB by itself)
            \ENDIF
        \ENDIF
    \ENDIF
\STATE Return (Compute, Network)
\end{algorithmic}
\label{NULB}
\end{algorithm}

In line with traditional VM scheduling algorithms promoting the best-fit packing, we investigate a variant of RISA, RISA-BF, prioritize boxes with lower available resources when INTRA\_RACK\_POOL is not empty. The main goal for RISA-BF is to better pack resources and reduce resource stranding. The resulting algorithm can be seen in Algorithm~\ref{RISA BF} and has been shown to have superior results when compared to the RISA.

\begin{algorithm}[ht!]
\caption{RISA-BF}
\begin{algorithmic}
\WHILE{$\exists$ unscheduled VM}
    \STATE Create $INTRA\_RACK\_POOL$
    \IF{$INTRA\_RACK\_POOL \neq \emptyset$}
        \STATE Sort boxes within each rack in ascending \# of resource
        \FORALL{$rack \in INTRA\_RACK\_POOL$ (round-robin)}
            \IF{$AVAIL\_INTRA\_RACK\_NET \neq \emptyset$}
                \STATE $Compute \leftarrow AllocCom(INTRA\_RACK\_POOL, REQ)$
                \STATE $Network \leftarrow AllocNet(AVAIL\_INTRA\_RACK\_NET)$
                \IF{$(Compute \neq \emptyset) and (Network \neq \emptyset)$}
                \STATE Assign VM
                \ENDIF
            \ENDIF
        \ENDFOR
    \ELSE
        \STATE Create $SUPER\_RACK$
    \STATE $Compute, Network \leftarrow  NULB(SUPER\_RACK, REQ, NET)$
        \IF{($Compute \neq \emptyset$) \AND ($Network \neq \emptyset$)}
            \STATE Assign VM
        \ENDIF
    \ENDIF
\ENDWHILE
\end{algorithmic}
\label{RISA BF}
\end{algorithm}

\subsection{Toy examples for comparative analysis}\label{toy_example}
In this section, we will see scenarios where RISA will outperform NULB and NALB. We will also see how RISA-BF may outperform RISA. 
\subsubsection{Toy example 1}
Firstly, let's consider Scenario 1. Table~\ref{scenario_1} lists the availability of compute resources. We are considering a typical VM with the following requirements - \textbf{8 cores of CPU, 16 GB of RAM, and 128 GB of storage}. Let us assume that there are enough network resources connecting these compute resources. 

\begin{table}[ht!]
\centering
\caption{Disaggregated architecture configuration for toy examples}
\begin{tabular}{|c|c|c|c|c|}
\hline
\multicolumn{5}{|c|}{CPU information}\\
\hline
id & rack & box & capacity & avail\\
\hline
0 & 0 & 0 & 64 cores & 0 cores \\
\hline
1 & 0 & 1 & 64 cores & 0 cores \\
\hline
2 & 1 & 0 & 64 cores & 64 cores \\
\hline
3 & 1 & 1 & 64 cores & 32 cores \\
\hline
\multicolumn{5}{|c|}{RAM information}\\
\hline
id & rack & box & capacity & avail\\
\hline
0 & 0 & 0 & 64 GB & 0 GB \\
\hline
1 & 0 & 1 & 64 GB & 16 GB \\
\hline
2 & 1 & 0 & 64 GB & 32 GB \\
\hline
3 & 1 & 1 & 64 GB & 16 GB \\
\hline
\multicolumn{5}{|c|}{Storage information}\\
\hline
id & rack & box & capacity & avail\\
\hline
0 & 0 & 0 & 512 GB & 0 GB \\
\hline
1 & 0 & 1 & 512 GB & 0 GB \\
\hline
2 & 1 & 0 & 512 GB & 256 GB \\
\hline
3 & 1 & 1 & 512 GB & 512 GB \\
\hline
\end{tabular}
\label{scenario_1}
\end{table}

Here, the CR for CPU is $0.08$, for RAM is $0.25$, and for storage is $0.17$. Hence, according to NULB or NALB, the CPU, RAM, and storage ids will be \textbf{(2, 1, 2)}. CPU and RAM need to communicate with each other. Also, RAM and storage need to communicate. In both of these cases, this will result in inter-rack network utilization. In comparison, with RISA, INTRA\_RACK\_POOL will now be equal to $[1]$. Thus, based on this, the VM will be assigned to CPU, RAM, and storage with ids \textbf{(2, 2, 2)}. In this case, there will be no additional inter-rack network utilization. Resulting in lower power utilization corresponding to this VM assignment. 

\subsubsection{Toy example 2}
In the second example, we will once again consider the DDC state in Table~\ref{scenario_1}. We will only consider the CPU requirements for subsequent VMs - \textbf{15 cores, 10 cores, 30 cores, 12 cores, 5 cores, 8 cores, 16 cores, and 4 cores}. Assuming, no previous VMs get released, let us see through Table~\ref{risa_vs_risa_bf} how RISA and RISA-BF will perform in this case. Considering all other compute and network resource requirements are met, based on the first-fit packing, box 0 first, and then box 1, RISA will continue to schedule VMs until id 5. It will drop the VM with id 6, due to inadequate resources and schedule VM with id 7. RISA-BF, however, by performing best-fit packing, alternating between box 0 and box 1, ends up allocating all of the subsequent VMs listed. In an actual scenario, this can either translate to a lower number of dropped VMs or fewer inter-rack VM assignments. Thereby, improving either resource utilization or network utilization. 

\begin{table}[ht!]
\centering
\caption{CPU requirement for subsequent VMs}
\begin{tabular}{|c|c|c|c|}
\hline
VM id & CPU req. & RISA Rack 1 box & RISA-BF Rack 1 box \\
\hline
0 & 15 & 0 & 1 \\
\hline
1 & 10 & 0 & 1 \\
\hline
2 & 30 & 0 & 0 \\
\hline
3 & 12 & 1 & 0 \\
\hline
4 & 5 & 1 & 1 \\
\hline
5 & 8 & 1 & 0 \\
\hline
6 & 16 & \textbf{NA} & 0 \\
\hline
7 & 4 & 1 & 0 \\
\hline
\end{tabular}
\label{risa_vs_risa_bf}
\end{table}

\section{Simulation results and comparative analysis}\label{sim_res}
\subsection{Performance evaluation using synthetic workload}
We will now see how NULB, NALB, RISA, and RISA-BF schedule a synthetic random workload on our disaggregated architecture discussed in Section~\ref{background}. We generated a random workload similar to the synthetic random workload in~\cite{Zervas}. A VM can have a random amount of CPU cores from 1 to 32 cores and a random amount of RAM from 1 to 32 GB. Storage for every VM is 128 GB. Requests are produced dynamically based on a Poisson distribution with a mean interarrival period of 10 time units. The VM life cycle begins at 6300 time units, with an increment of 360-time units for each set of 100 requests. A total of 2500 VMs were generated.

\begin{figure}[ht]
    \centering
    \includegraphics[width=\linewidth]{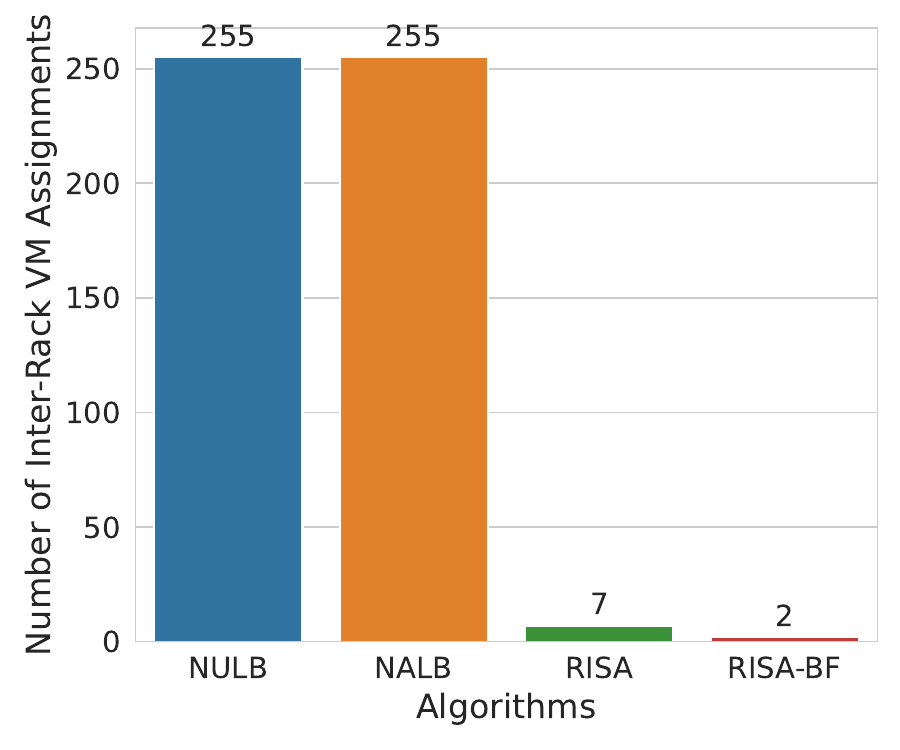} 
    \caption{Number of inter-rack VM assignments}
    \label{syn_res}
\end{figure}

The average CPU utilization for all algorithms was $64.66\%$, the average RAM utilization for algorithms was $65.11\%$ and the average storage utilization for algorithms was $31.72\%$. As seen in Figure~\ref{syn_res}, despite utilizations well below $100\%$, both NULB and NALB had 255 inter-rack VM assignments. On the other hand, RISA and RISA-BF had only 7 and 2 inter-rack VM assignments respectively. 

\noindent
In the next subsection, we will see how this difference in inter-rack VM assignment translates into substantially lower power consumption (for optical components)for RISA and RISA-BF, in comparison to NULB and NALB, for a practical workload. Similarly, we will also see how it translates into a lower average CPU-RAM latency for RISA and RISA-BF in comparison to NULB and NALB.

\subsection{Performance evaluation using practical workload}
We will now compare the performance of the algorithms in the areas of network utilization, energy consumption, and average CPU-RAM latency using the 2017 public release of Microsoft Azure data center traces~\cite{azure2023}. To choose workloads of different characteristics, we selected different subsets of the 2017 Azure data center traces - the first 3000 VMs~(Azure-3000), the first 5000 VMs~(Azure-5000), and the first 7500 VMs~(Azure-7500).

\begin{figure}[ht]
    \centering
    
    \begin{subfigure}[b]{\linewidth}
        \centering
        \includegraphics[width=\textwidth, height=4cm]{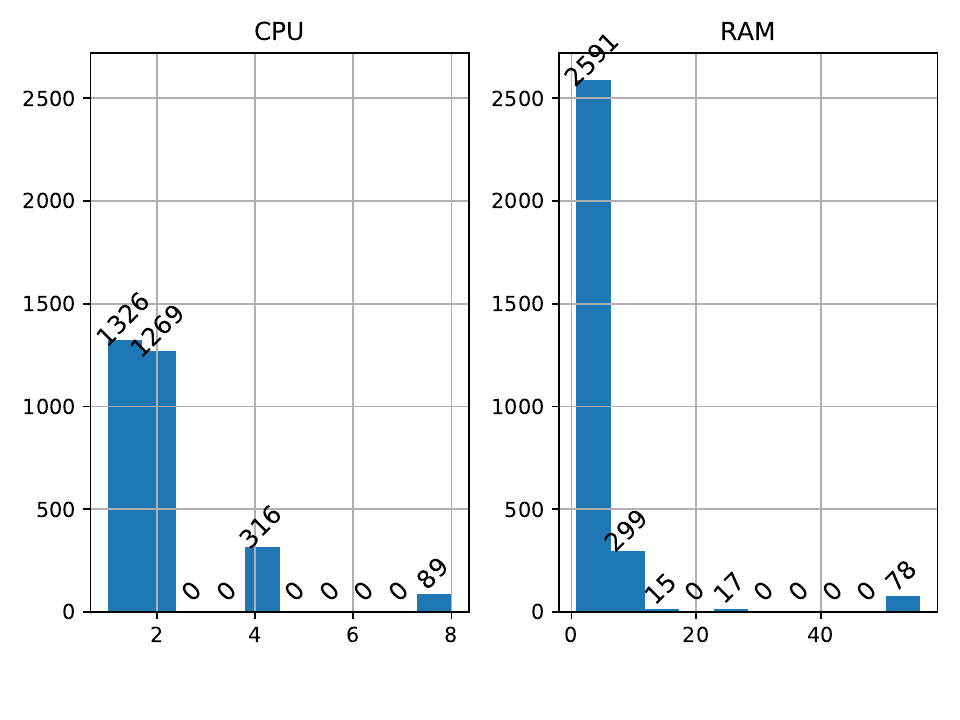}
        \caption{Azure-3000}
    \end{subfigure}

    \begin{subfigure}[b]{\linewidth}
        \centering
        \includegraphics[width=\textwidth, height=4cm]{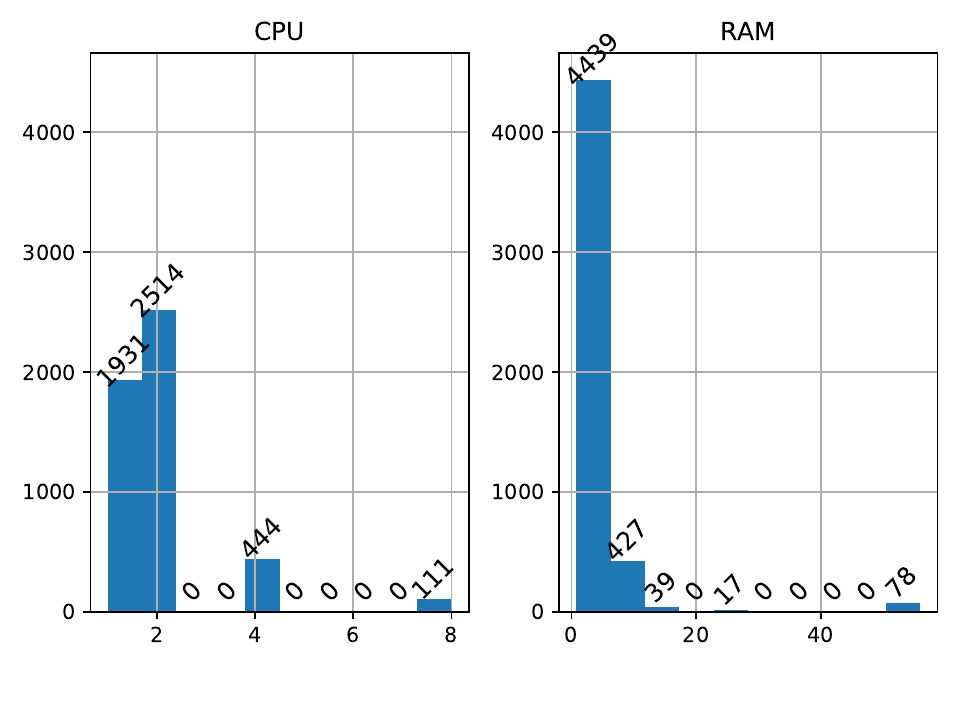}
        \caption{Azure-5000}
    \end{subfigure}

    \begin{subfigure}[b]{\linewidth}
        \centering
        \includegraphics[width=\textwidth, height=4cm]{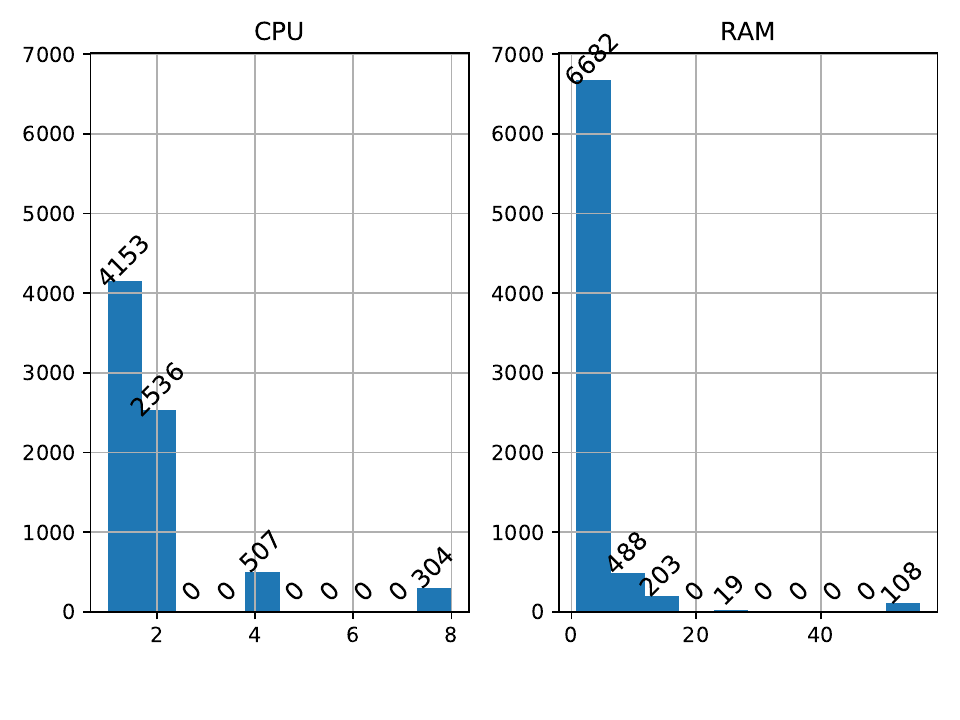}
        \caption{Azure-7500}
    \end{subfigure}
    
    \caption{CPU and RAM distribution of Azure data center traces}
    \label{workload_char}
\end{figure}

There are many ways in which the Azure data center workload is different from the random workload. Firstly, the CPU requirement is generally low. In comparison to the CPU requirement, the RAM requirement for some VMs is quite high. For this workload, we assume storage to be 128 GB, similar to~\cite{Zervas}. Thus, in most cases, storage is the most contended resource. In Figure~\ref{workload_char}, we see the characterization of the different practical workloads in terms of their CPU and RAM requirements. From Figure~\ref{workload_char}, it is seen that Azure-3000 has the lowest percentage of small VMs compared to Azure-5000 and Azure-7500.  For Azure-5000, the percentage of smaller VMs is more. Azure-7500 has the greatest percentage of small VMs. Thus, the scheduling scenarios for each of these workloads are different. The network requirement for the VMs can be obtained based on Tables~\ref{conf} and \ref{net_req}. We will see how the RISA and RISA-BF perform in the scheduling task in comparison to NULB and NALB.


\begin{figure}[ht]
    \centering
    \includegraphics[width=\linewidth]{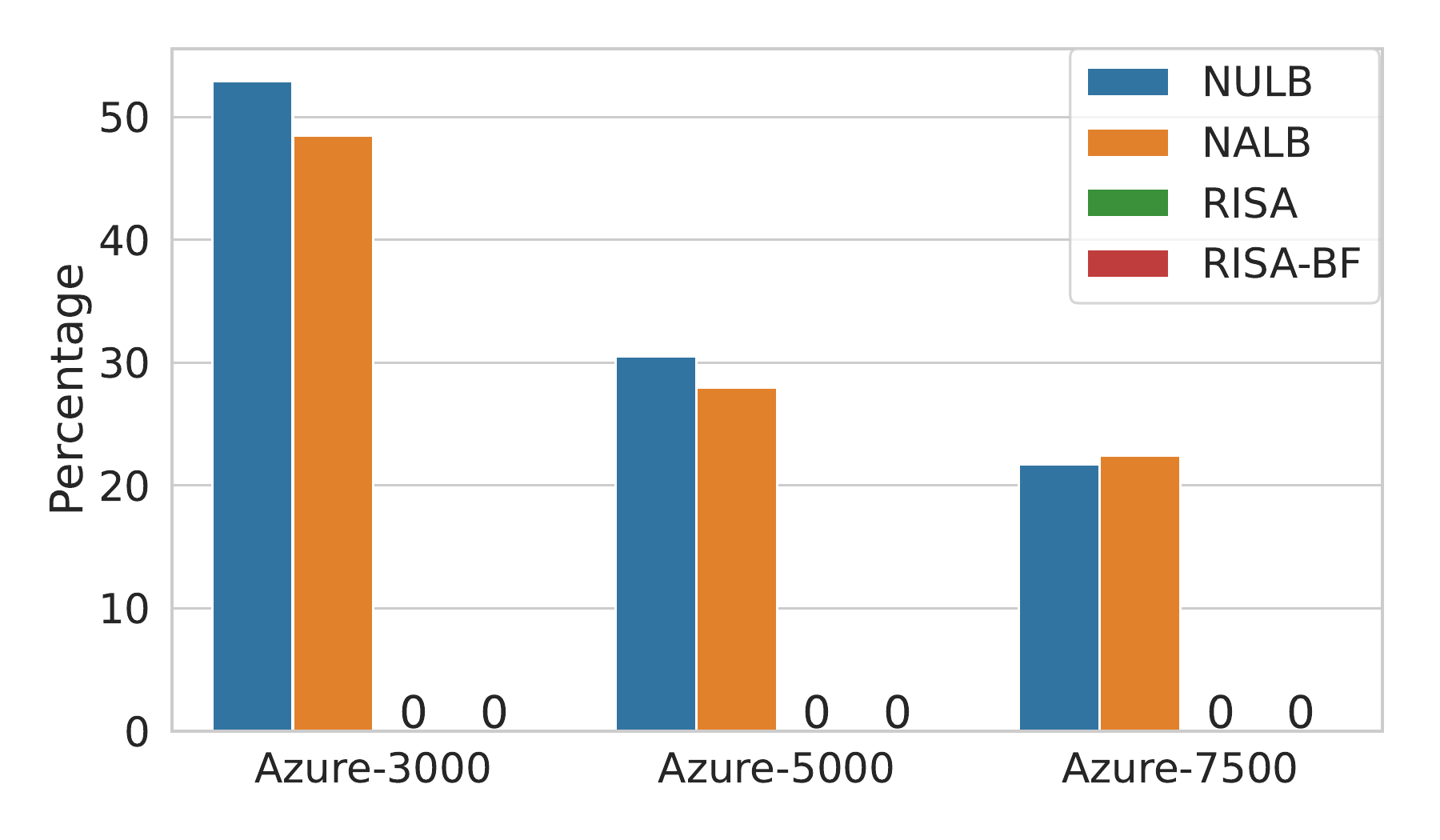} 
    \caption{Percentage of inter-rack VM assignments}
    \label{inter_rack_VM}
\end{figure}

Figure~\ref{inter_rack_VM} shows the percentage of inter-rack VM assignments (out of the total number of VMs) for the three types of workloads discussed in the previous section. Here, we can see that both NULB and NALB have significant amounts of inter-rack assignments, up to $52\%$ and $48\%$ for NULB and NALB, respectively. Notably, RISA and RISA-BF have no inter-rack VM assignments for any of the workloads - showing their ability to exploit intra-rack VM allocations while leaving room for future VM requests.

\begin{figure}[ht]
    \centering
    \includegraphics[width=\linewidth]{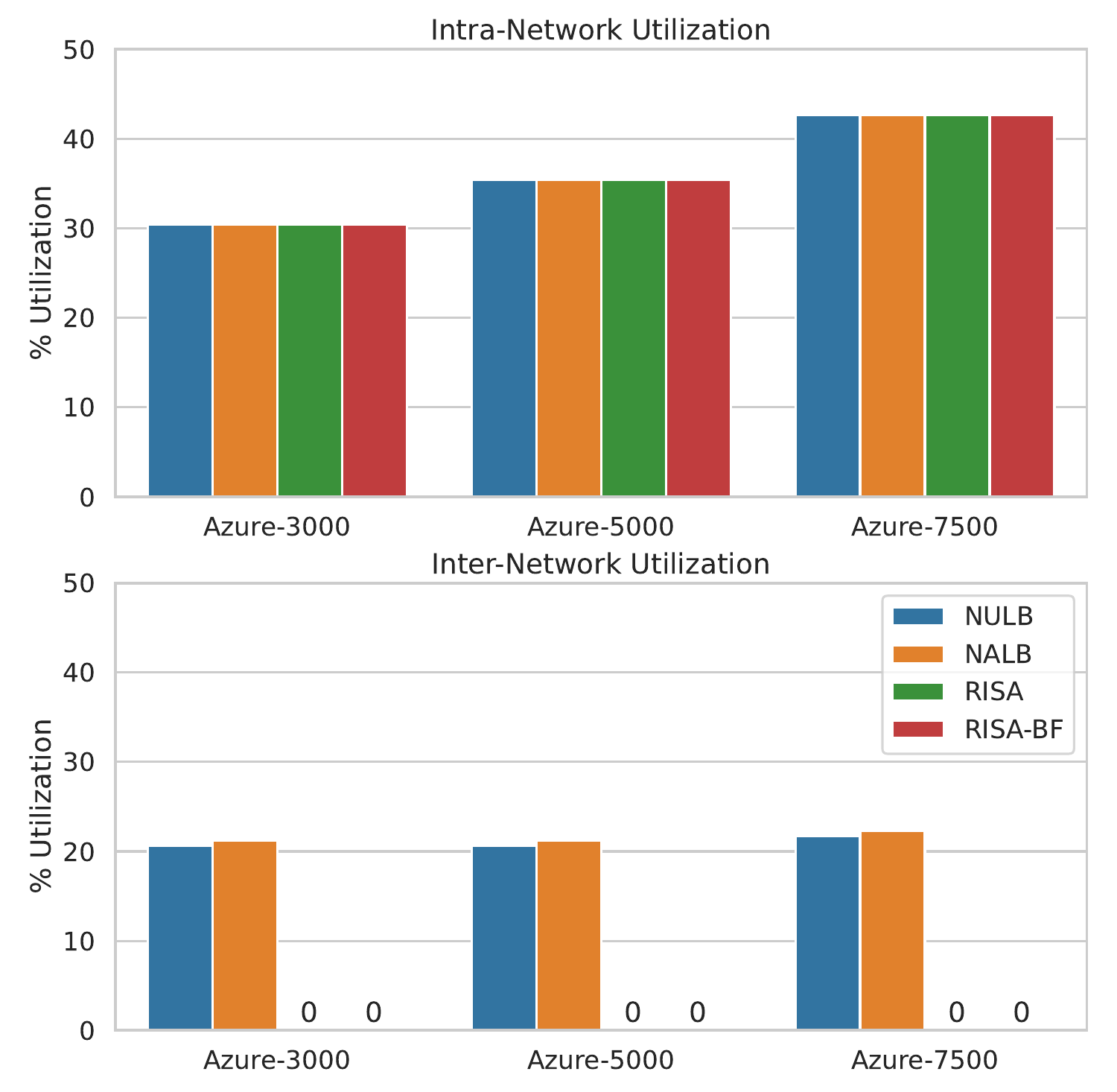} 
    \caption{Network utilization}
    \label{net_util}
\end{figure}

For the DDC, each box only contains one type of resource, all VMs have to use the rack switch to communicate between resources regardless of where the resources reside. Thus, when the same amount of resources is used, an equal amount of intra-rack bandwidth is used to access the rack switches. For the scheduling problems in the discussion, no VMs were dropped during the scheduling process. Thus, accordingly, as seen in Figure~\ref{net_util}, the intra-rack network utilization of all the algorithms are the same - $30.4\%$ for Azure-3000, $35.4\%$ for Azure-5000, and $42.6\%$ for Azure 7500. However, the inter-rack network utilization for all the algorithms are not the same. As explained in Section~\ref{background}, RISA and RISA-BF try to minimize the amount of inter-rack VM assignments. This results in fewer inter-rack resources being used. Subsequently, resulting in lower power consumption from optical components.  For the three workloads in the discussion, since there are no inter-rack VM assignments for RISA and RISA and RISA-BF, it can be seen from Figure~\ref{net_util} that the inter-rack network utilization is also zero.

\begin{figure}[ht]
    \centering
    \includegraphics[width=\linewidth]{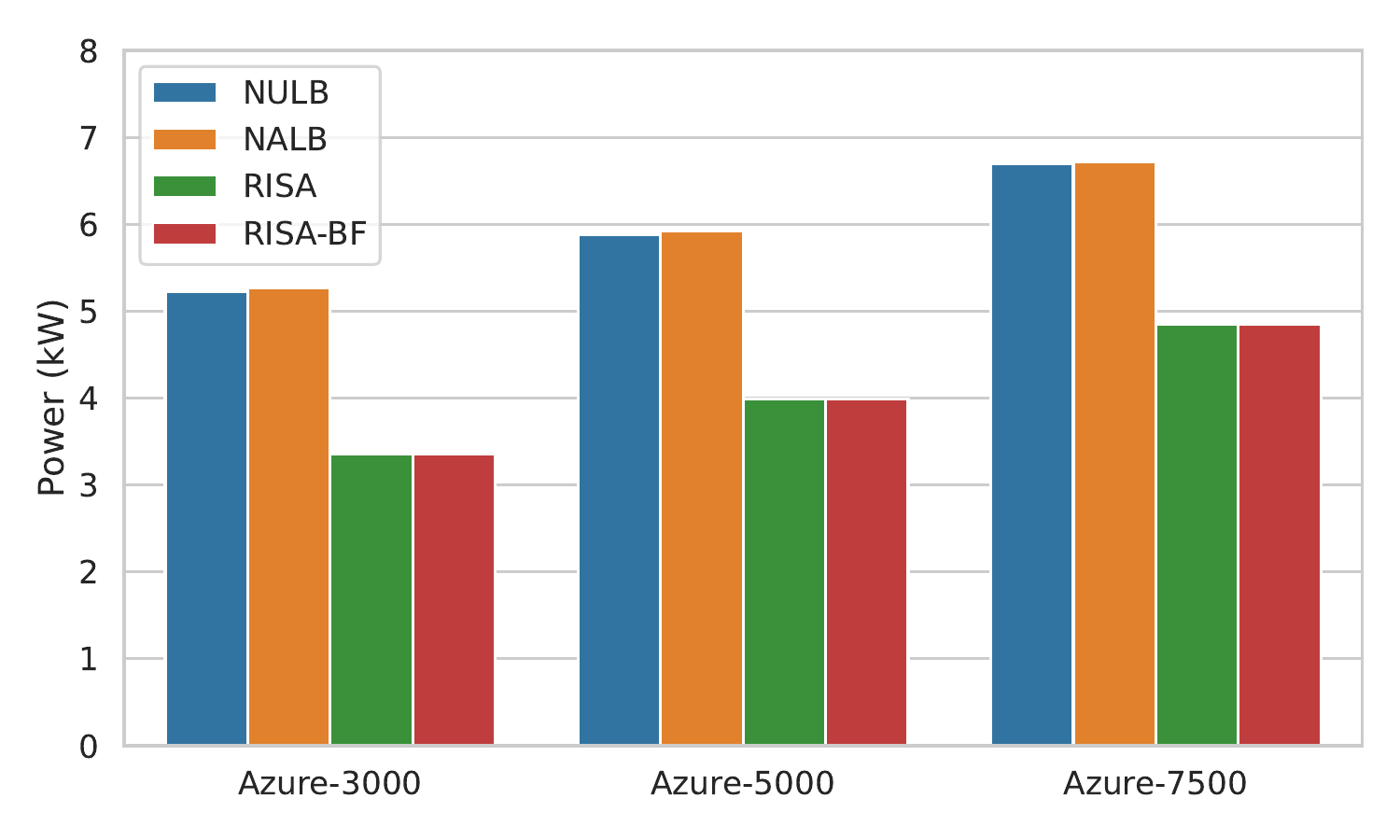} 
    \caption{Power consumption for optical components}
    \label{opt_power}
\end{figure}

Inter-rack switches usually have much higher port numbers in comparison to rack or box switches. This is because they connect to a large number of racks within a cluster. Thus, if an algorithm requires more inter-network bandwidth, it ends up consuming a larger amount of power from optical components, particularly from the inter-rack switches. For the power consumption calculation for optical components, we considered the transceiver power and total optical switch (box switch, intra-rack switch and inter-rack switch) power. To support the DDC architecture in Section~\ref{background}, we need to have box switches with 64 ports, intra-rack switches with 256 ports and inter-rack switches with 512 ports. In Figure \ref{opt_power}, for Azure-7500, the power consumption due to optical components is seen to be as high as $6.70$ kW and $6.72$ kW for NULB and NALB respectively. Also, it can be seen that the same power consumption for RISA and RISA-BF can be as low as $3.36$ kW for Azure-3000, whereas for NULB and NALB, the values were $5.22$ kW and $5.27$ kW respectively. This shows that RISA or RISA-BF each have $33\%$ reduction in power consumption from optical components, as compared to NULB or NALB.

\begin{figure}[ht]
    \centering
    \includegraphics[width=\linewidth]{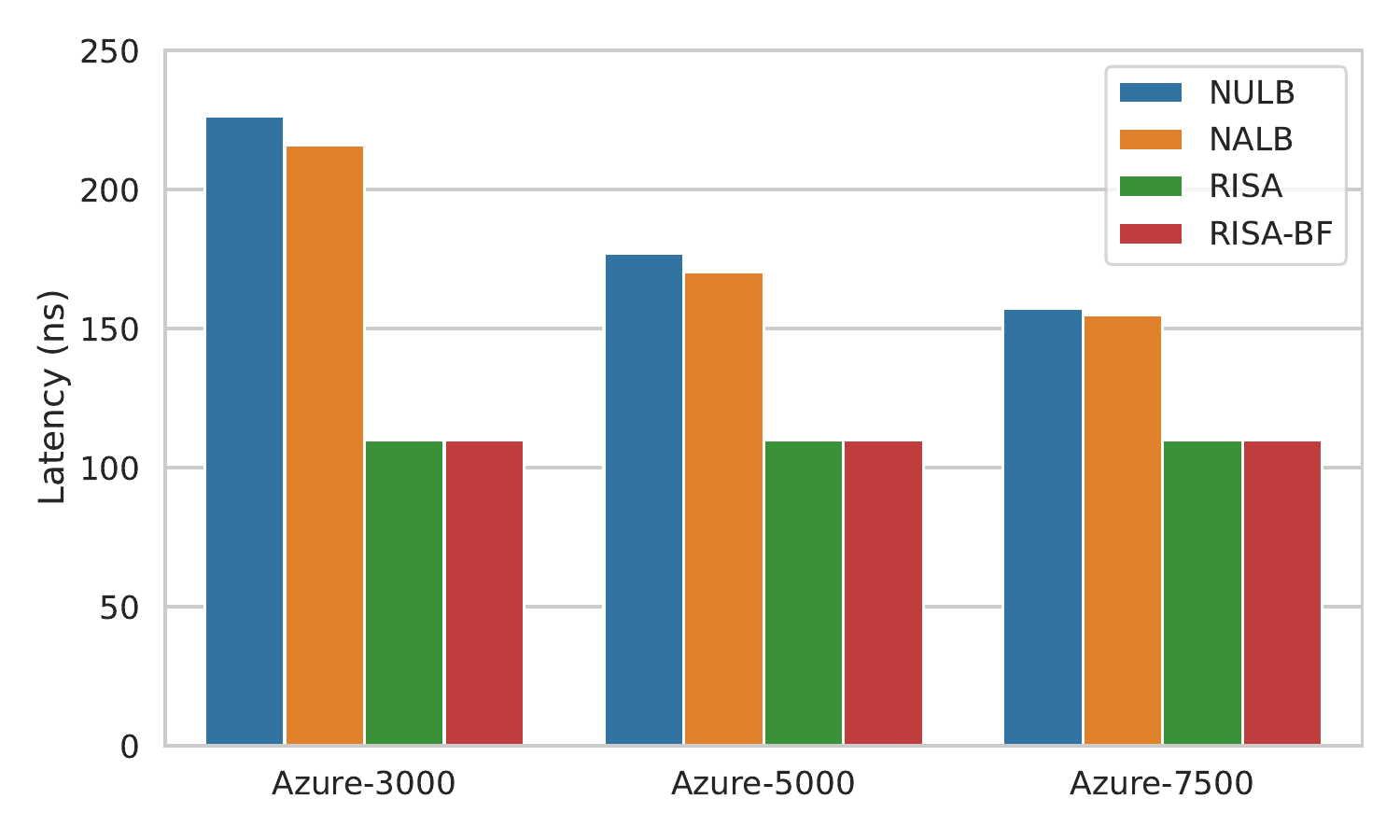} 
    \caption{Average CPU-RAM delay}
    \label{delay}
\end{figure}

Quality of service (QoS) can be an important aspect when it comes to scheduling of VMs within a DC. Since data centers sometimes run third-party workloads, QoS becomes even more of a concern, which cannot change with a change in workload. Thus, the last area of focus in comparing the three algorithms is the average CPU-RAM round-trip latency. From~\cite{Zervas}, we assume that there is a $110$ ns CPU-RAM round-trip latency for communication within a rack, and that across racks is $330$ ns. From Figure~\ref{delay}, it can be seen that for Azure-3000, NULB and NALB have an average CPU-RAM round-trip latency of $226$ ns and $216$ ns respectively. For RISA or RISA-BF, it is only $110 ns$. This shows that for Azure-3000, RISA or RISA-BF has lower than half of the average CPU-RAM round-trip latency as compared to NULB or NALB. Thus, RISA and RISA-BF are significantly better when QoS is of higher priority. For inter-rack center switches with a larger number of ports, the inter-rack delay may be higher, so the values in Figure~\ref{opt_power} represent optimistic values for average CPU-RAM round-trip latency. However, since RISA and RISA-BF both out-perform NULB and NALB in terms of inter-rack VM allocations, we expect RISA and RISA-BF to have even larger improvements in CPU-RAM latency for larger systems.

\subsection{Execution times of different algorithms}
For RISA or RISA-BF, when $INTRA\_RACK\_POOL$ is empty, it uses NULB for finding required compute and network resources for VM assignment. Thus, the time complexity for NULB, RISA and RISA-BF are the same. NALB performs an additional step of finding the path with the most available network bandwidth. Thus, the time complexity of NALB is higher compared to NULB, RISA, and RISA-BF. However, in practice, $INTRA\_RACK\_POOL$ is not always empty. In fact for the simulation results discussed in preceding subsections, $INTRA\_RACK\_POOL$ was never empty. Thus, in all cases, scheduling under RISA ended first. After RISA, scheduling under RISA-BF ended second, scheduling under NULB ended third, and scheduling under NALB ended last. 

\begin{figure}[ht]
    \centering
    \includegraphics[width=\linewidth]{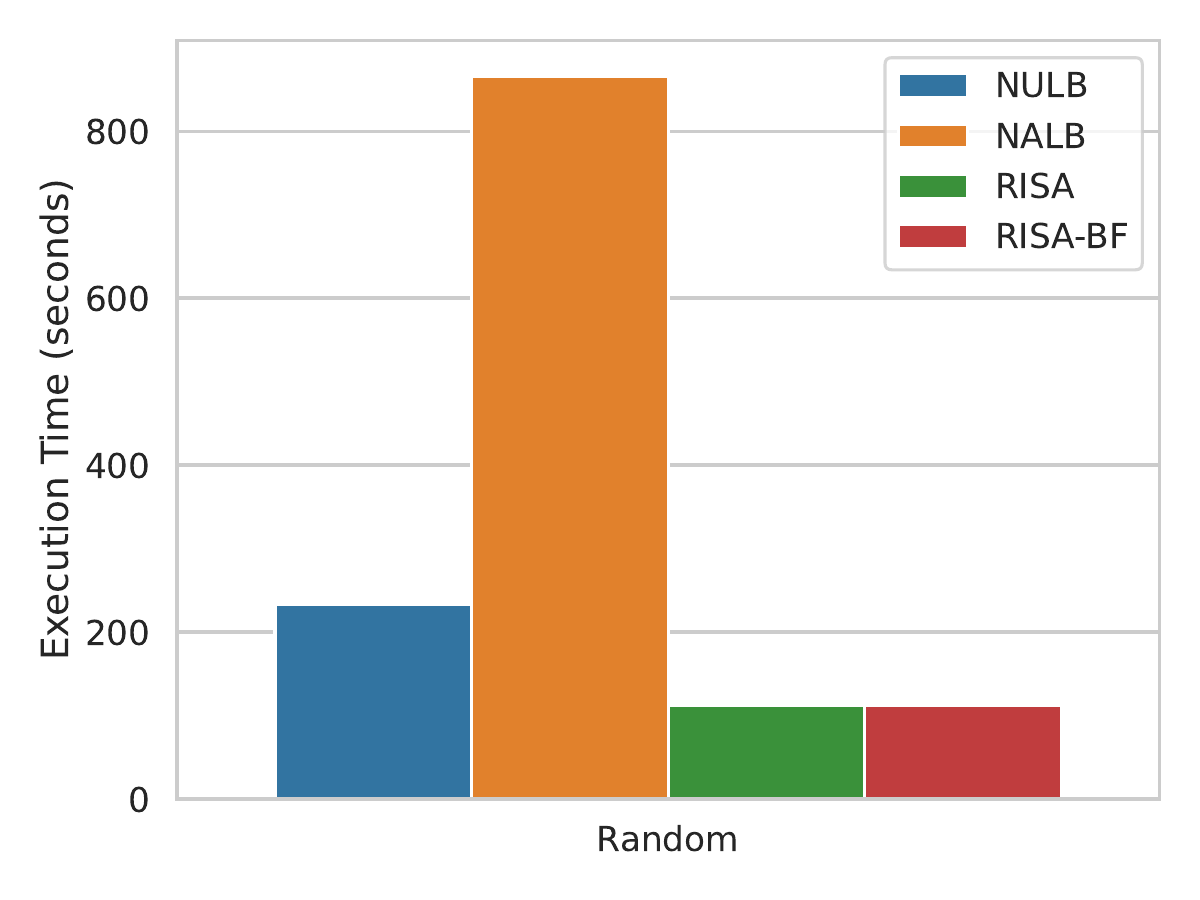} 
    \caption{Execution time of synthetic workload}
    \label{exec_syn}
\end{figure}

\begin{figure}[ht]
    \centering
    \includegraphics[width=\linewidth]{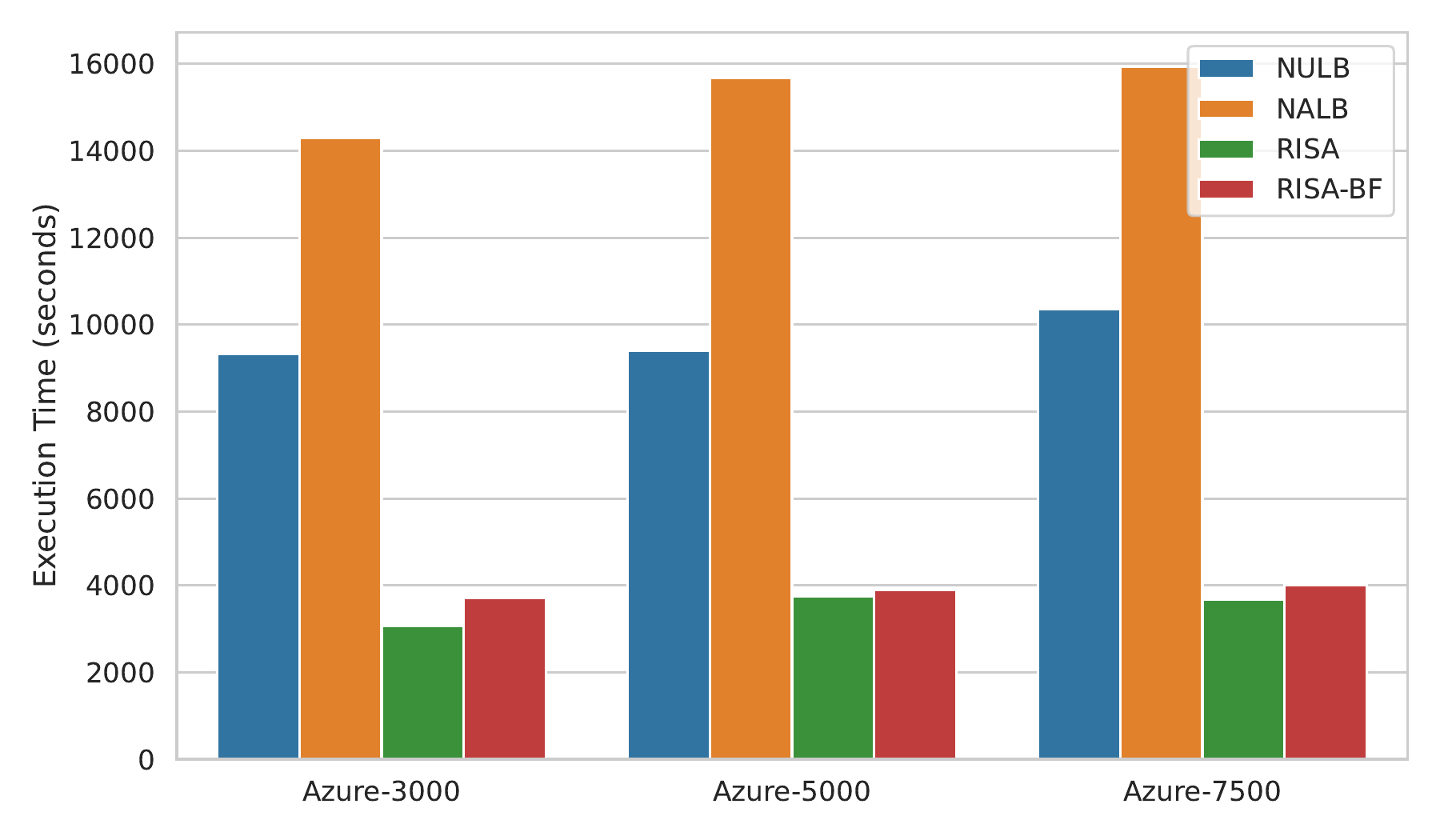} 
    \caption{Execution time of practical workload}
    \label{exec_prac}
\end{figure}

Table~\ref{system_config} lists the configuration of the system, which was used to run the simulations. Figure~\ref{exec_syn} gives a visual representation of the scheduling times of the algorithms for the synthetic workload. It can be seen that the execution time of NALB is the highest, at $865 seconds$, which is followed by NULB, at $233$ seconds. RISA takes $111$ seconds and RISA-BF takes $112$ seconds. For RISA or RISA-BF this translates to greater than 2 $\times$ speedup when compared to NULB and close to 8 $\times$ speedup when compared to NALB. Figure~\ref{exec_prac} gives a visual representation of the scheduling times of the algorithms for the practical workload. Here, for Azure-7500, the execution time for NALB was $15929$ seconds and that for NULB was $10361$ seconds. For RISA and RISA-BF these values were  $3679$ seconds and $4013$ seconds respectively. Thus, for RISA this translates to 2.81 $\times$ speedup when compared to the execution time of NULB. For RISA, the speedup when compared to the execution time of NALB is 4.33$\times$. 

\begin{table}[ht!]
\centering
\small
\begin{tabular}{|l|l|}
\hline
\textbf{Component} & \textbf{Specification} \\
\hline
Processor & AMD Ryzen 7 2700X, 4.3 GHz (8 cores, 16 threads) \\
\hline
RAM & 4$\times$8GB\,DDR4, 2133 MT/s\\
\hline
\end{tabular}
\caption{Simulation System Configuration}
\label{system_config}
\end{table}

\section{Conclusion}\label{conc}
The goal of this paper was to propose an approach to significantly reduce the network utilization, power consumption, and CPU-RAM round-trip latency. We have been successful to reduce the network utilization significantly, which resulted in up to more than $33\%$ reduction in power consumption of optical components. Compared to the state-of-the-art, our approach achieved up to $50\%$ reduction in CPU-RAM round-trip latency. Additionally, for practical workload, we achieved up to 2.81 $\times$ speedup when compared to NULB, and up to 4.33$\times$ speedup when compared to NALB. 

\begin{acks}
This work was supported by the National Science Foundation (NSF) under grant number CNS-2046226.
\end{acks}

\bibliographystyle{ACM-Reference-Format}
\bibliography{sample-base}

\appendix

\end{document}